# Portfolio optimization using local linear regression ensembles in Rapid Miner

Gábor Nagy, Tamás Henk PhD.,
Budapest University of Technology and Economics

**Abstract**

In this paper we implement a Local Linear Regression Ensemble Committee (LOLREC) to predict 1-day-ahead returns of 453 assets form the S&P500. The estimates and the historical returns of the committees are used to compute the weights of the portfolio from the 453 stock. The proposed method outperforms benchmark portfolio selection strategies that optimize the growth rate of the capital. We investigate the effect of algorithm parameter m: the number of selected stocks on achieved average annual yields. Results suggest the algorithm's practical usefulness in everyday trading.

## 1.   Introduction

In this paper we present a sequential investment strategy – a portfolio selection strategy or portfolio optimization technique – that could be used in financial markets. Sequential investment means that at the end of one trading period the investor is allowed to redistribute his current capital among a set of available assets. The investor's goal is to maximize his capital. The portfolio selection is based on historical data collected from the market. Local linear regression base models or experts are used in an ensemble called a committee to model the next-day return of an asset. The committees use different voting strategies to provide the estimate for each asset. The estimates along with historical performances will be used to generate portfolio weights for a given trading period. Numerical results will be presented to show the performance of the portfolio selection strategy.

### 1.1.   Stock market model

Our model of the stock market follows the general model presented in [4]. The stock market is comprised of d assets. A market vector $X = (x_1, x_2,\ldots, x_d)$ where $x_j \geq 0$ is the price relative of the given trading period that represents the growth of the capital invested in the $j^{th}$ asset. Diversification of capital is achieved by a portfolio vector $B = (b_1, b_2, \ldots, b_d)$. Here $b_j \geq 0$ represent the portion of capital invested in the $j^{th}$ asset. The portfolio vector $B$ is constructed such that $\sum b_j = 1$. This means the strategy is self-financing and no withdrawal of capital is allowed, furthermore because of the non-negativity of $b_i$ no short selling or buying on margin is allowed. The stream of market vectors $X_1, X_2, \ldots, X_i$ represent the evolution of the market and for all $X_i$ the investor selects a portfolio vector $B_i$. The achieved wealth in each market period is computed by Formula 1.

$$\begin{aligned}
S_0 &= 1 \\
S_1 &= S_0 X_1 B_1^T \\
S_2 &= S_1 X_2 B_2^T \\
&\ldots \\
S_i &= S_{i-1} X_i B_i^T
\end{aligned} \quad \text{(Formula 1)}$$

So in the $0^{th}$ period we start trading with 1 unit of hypothetical dollars. The following period's wealth can be computed by the product of the previously accumulated wealth, the new market vector and the portfolio vector.

## 1.2. Portfolio weight estimation with ensemble methods

There are two approaches modeling the evolution of the market: allowing $X_i$ to take arbitrary values without a stochastic model [5,6] or assuming that the prices are realizations of a random process and describe a statistical model [2,3]. The former papers shows that a finite number of base models called experts infer the unknown distribution of the underlying random process that generates the market vectors. The experts than generate a portfolio vector that maximizes the wealth given the empirical distribution of the data collected. Experts are combined weighed by their past performance (the wealth achieved so far) generating a portfolio vector, that maximizes the growth rate of the capital.

Our approach uses historical returns to estimate next-period returns of individual assets in the portfolio with local linear regression base estimators of one-period-ahead returns called experts that are set with different parameters. We use the experts to create a committee with different voting functions to aggregate the base predictions of the individual experts.

This technique is referred to as ensemble methods in machine learning. [10] Numerous papers suggest that ensembles combining many simple base models perform better and provide greater accuracy compared to only one highly sophisticated model, that have been tuned extensively. [11,12] This ensemble method is used in random forests where a vast number of trees are built on different attribute subsets and averaged to provide prediction in classification. [1] This technique show up in the technique presented in [2,3] and we will also use this approach in our algorithm in the committees that estimate the next-period return using base predictors with different parameters. We want to emphasise that one could use other modeling techniques for next-period return estimation, or use an ensemble of ensembles. We will discuss these possibilities in the last section.

Győrfy mentions that if d is large there are not enough historical data to infer the distribution of the market vector. [2] By simplifying the problem to individual assets overcomes this problem of d being a large number and overcome the problem of "curse of dimensionality".

Furthermore methods using the entire market vector has a problem that if one wants to perform modeling on a different set of assets one have to construct the model all over again, which is very time consuming. By individual asset return estimation we may lose information that is represented in cross-asset dependencies, thus generate lower growth rate of the capital, but gain an advantage in decreased runtime of portfolio vector estimation. Furthermore each expert's output estimation and the committee estimates can be reused: they only need to be generated once. This way we can perform analysis of multiple portfolios easily considering a universe of all possible assets. A previously computed portfolio's subset can be analyzed even faster since only the weights have to be recalculated.

The proposed method focuses more on practical usefulness, ease of implementation, however numerical results will show that our approach outperforms highly optimized portfolio selection strategies on reference datasets.

## 2. Implementation

### 2.1. Local linear regression experts

The section explains how one local linear regression base expert is built, what input data is used and how the learning scheme is applied. In general a base model in an ensemble could be any model we have chosen local linear regression because recent studies suggest the practical usefulness of the method in return forecasting [7-10].

Local Polynomial Regression in RapidMiner takes a number of parameters; by setting the degree parameter to 1 we get a Local Linear Regression. We refer to parameters that are not set in our algorithm as *constant* and *variable* if we allow setting of those parameters. Table 1 show these parameters along with a window size parameter that sets the lookback window size $w$ of a Windowing operator that transforms the single column input market vector for an arbitrary asset into an ExampleSet containing past return data. We will use the notion for an expert $E_j(k,w)$: the expert is operating on the $j^{th}$ asset with, the $k$ closest neighbors are selected to be used in the local linear regression and the window transformation with window size $w$ will be performed before the learning is started.

| degree | 1 | constant |
|---|---|---|
| ridge | 0.01 | constant |
| numerical distance measure | Eucledian | constant |
| neighbourhood type | fixed number | constant |
| k | variable ||

| smoothing kernel | Exponential | constant |
|---|---|---|
| window size (w) | variable ||

Table 1. Parameters of local linear regression implementation in RapidMiner

We show a general dataset that is used to build one expert with *w=3* in Table 2. The *label* variable is the label (target), *predictor-1, predictor-2* and *predictor-3* is the predictor variables that are generated by the Windowing operator[1]. A sliding window is used to extend the train set size of the model. The *train_set* attribute indicates that an example is used in the training phase: A value of 1 indicates it is in the training set, 2 indicates it is in the application set. In our setting the train set starts from the first example and lasts with the *i-1*$^{th}$ example the model is applied on the *i*$^{th}$ example in the *i*$^{th}$ iteration. Note that this learning scheme does not use validation in any form.

| label | predictor-1 | predictor-2 | predictor-3 | train_set |
|---|---|---|---|---|
| x3 | x2 | x1 | x0 | 1 |
| x4 | x3 | x2 | x1 | 1 |
| ... | | | | 1 |
| xi-1 | xi-2 | xi-3 | xi-4 | 1 |
| xi | xi-1 | xi-2 | xi-3 | 2 |
| ... | | | | NA |
| xn | xn-1 | xn-2 | xn-3 | NA |

Table 2. A general outline of an example set generated from an n length market vector of an arbitrary asset used to train a local linear regression model

One could use a filter parameter to reduce the size of the training set to speed up the model building and exclude examples from the past. Past performance – for example the return achieved by the expert – could be used to weigh an expert in the committee, however neither previously mentioned improvements had been implemented in the current to setup.

## 2.2. Local linear regression committee

The committee for the *j*$^{th}$ asset $C_j(K,W,V)$ consist of local linear regression experts $E_j(k,w)$ where $k \in K$, $K \in Z^+$, $w \in W$, $W \in Z^+$ and the committee's voting function is *V*. In the *i*$^{th}$ iteration let the estimate of $E_j(k,w)$ be $e_{ij}$ forming a vector of estimates $e_i$ of *j* components. The voting function *V* is an arbitrary function where

---

[1] For practicality reasons we generate a larger window size and select the predictors using regular expressions

$v_{i,j} = V(e_i, p_1, p_2, \ldots, p_m)$ where $p_1, p_2, \ldots, p_m$ are optional parameters that could be set – as in the previous section – these could be past performances or precomputed weights. The resulting $v_{i,j}$ is the committee's output for the $i^{th}$ iteration using a voting function $V$ operating on the $j^{th}$ asset. The current implementation does not use parameters in the voting function just the estimates. Three different voting functions are used:

1. **Average voting function**: That weighs the base experts evenly. By $|C_j(K,W,V)|$ we refer to the number of experts used by the committee.

$$v_i = \frac{1}{|C_j(K,W,V)|} \sum e_i \qquad \text{(Formula 2)}$$

2. **Median voting function**: The median of the base expert estimates.
3. **Mode voting function**: The mode is the most probable estimate based on the distribution of the estimates of the base experts.

It is important to note that one can set other parameters such as the smoothing kernel or distance function of the base local linear regression experts or set more K, W values resulting in more experts. This decision is essentially a function of computational power at hand or time that one wishes to take to perform the predictive step. In later sections it will be shown that the increase in the number of committee members increases the overall yield of the algorithm, however also increases the computational time needed to perform the predictive step.

Analog to the base experts, a committee also has a performance $S_{i-1}^{C_j(K,W,V)}$ at the $i^{th}$ iteration: this is the accumulated wealth of the committee working on the $j^{th}$ asset in the $i\text{-}1^{th}$ iteration (since we do not know the return of the $i_{th}$ trading period we cannot use $S_i^{C_j}$). We will use this performance measure as a weight of the estimate $v_{i,j}$ of the committee in generating portfolio weights.

### 2.3. Generating portfolio weights

The selection of a portfolio is the method, algorithm or strategy one finds values for $B_i$. Dynamic asset allocation allows investors to compute $B_i$ for each trading period. Portfolio weight calculation in our setup is based on a heuristic: no optimization is done on historical data, only the committee estimates and the committee's accumulated wealth $S_{i-1}^{C_j(K,W,V)}$ are used.

The main reason behind this is practicle: runtime optimization. If one wants to experiment with dozens of assets on a long timeframe runtime will become an issue as one model may be relevant theoretically but could not be implemented because of runtime limitations: the weights must be computed in each trading period.[2] If the trading period length is long (eg. 1 week) this is not an issue, however if the granularity is small (1 hour to a few minutes) – as it is often the case with automated trading strategies – one may find theoretically optimal models impossible to implement as extensive portfolio weight optimization with current technology is impossible given the frequency one have to recalculate the weights.

After create the universe of stocks by gathering data with the given granularity the Local Linear Regression Committee (LOLREC) portfolio weight estimation method can be split into 7 steps. We present the algorithm in the $i^{th}$ iteration. The first iteration LOLREC uses an equally weighted portfolio vector.

1. Build $C_{i,j}(K,W,V)$ described in Section 2.2 for each asset.
2. Compute $v_{i,j}$ using committee $C_{i,j}(K,W,V)$ using the base estimates $e_{i,j}$ of experts $E_j(K,W)$ that are formed based on the dataset from the 1$^{st}$ row to the $i^{th}$ row.
3. Set $v_{i,j} = 0$ where $v_{i,j} < 1.0$.
4. Let $v_{i,1}, v_{i,2}, ..., v_{i,m}$ be the m largest asset return estimates.
5. Compute the $i^{th}$ portfolio vector $B_i = (b_{i,1}, b_{i,2}, ..., b_{i,d})$ with Formula 2.

$$b_{i,j} = \frac{e_{i,j} * S_{i-1}^{C_{i,j}(K,W,V)}}{\sum_j b_{i,j}} \qquad \text{(Formula 2)}$$

The $j^{th}$ component of the portfolio vector is the product of the committee's estimate and committee's historical performance, normalized by the sum of the vector components. Note that in the implementation we first compute the $b_{i,j}$ using the denominator and normalize only if $\sum_j b_{i,j}$ is not 0.0, that is all $v_{i,j} < 1.0$, thus truncated to 0.0. In this occasion capital rests in cash with return 1.0, because no return estimate offers an increase in capital.

6. The return of the portfolio and the accumulated wealth of the individual committees are calculated with Formula 3.

$$S_i = S_{i-1} X_i B_i^T \qquad \text{(Formula 3)}$$

7. The committees return is calculated by Formula 4.

$$S_i^{Cj(K,W,V)} = S_{i-1,j} I_{i,j}$$
$$I_{i,j} = \begin{cases} X_{i,j} & \text{if } e_{i,j} > 0 \\ 1 & \text{otherwise} \end{cases} \qquad \text{(Formula 4)}$$

---

[2] 5 years of FOREX data with a granularity of 30-minute-long trading periods comprise of 87600 rows

In the next section we show some numerical results that the strategy achieves on reference datasets and recent data. The only super-parameter of the LOLREC – disregarding the vast abilities one could experiment with smoothing kernels, distance functions and different K, W values in the base experts – is *m*, that is the number of assets with largest estimated next-period returns to be selected.

## 3. Datasets

Three datasets were used to investigate the efficiency of the LLRE-PWE algorithm. We will refer to them as: SNP500, NYSEMERGED and NYSEOLD. The former two are benchmark datasets used in [2,3]. We will compare our results to theirs where applicable.

SNP500 consists of 453 different stock's daily returns. Stocks were chosen from stocks of that makes up the S&P500 index. The dataset ranges from 17.04.2007 to 17.04.2012 and gives 5 years of data in 1260 examples. Stocks used had no missing values (there were price movements for all days in the given interval) and we excluded stocks that were either removed or added to the index, gone bankrupt or bought by a third party during the investigation period. This introduces a selection bias in our results. The returns were computed from split and dividend adjusted closing prices.

The NYSEMERGED dataset have 19 stock's daily returns with a 44 year timespan (11178 trading days that end in 2006). This dataset were used in a number of articles, see [2,3]. The stocks in this dataset are from the New York Stock Exchange.

NYSEOLD consist only 22 years and 33 assets daily returns (5651 trading days ending in 1985). Some of the assets here can also be found in the NYSEMERGED dataset (those that did not go bankrupt the time after NYSEOLD dataset had ended). This dataset is also used in various papers [2,3].

## 4. Results

### 4.1. Results of LOLREC on NYSEOLD dataset

First we present the numerical results on the benchmark dataset NYSEOLD. For the experts $C_j(K,W,V)$ is set with K=[1,…,10], W=[1,…,5] for each asset. All three different voting functions were tested: the average, median and mode voting function. The LOLREC is called with m=10 for selecting the top 10 predictions from all 33 assets.

The largest yield at the end was achieved by the committees with average voting strategy. If we would have invested $1 into this portfolio selection scheme in 1963 on 1985 we could have had $5.09*10$^9$. The best benchmark – that is a kernel based log-optimal dynamic portfolio selection strategy reported in [2] – achieved $5.63*10^8$ in the end. The average annual yield[3] (AAY) shows, that the best performing voting strategy had AAY of 276%. The lowest AAY and was generated by the mode voting strategy, that resulted in a lower AAY that of the benchmark. The results are shown in Table 3 coupled with the trading period.

| Period | C2 | | | Benchmark |
|---|---|---|---|---|
| | average | median | mode | BK(1.0) |
| 500 | 7.73E+00 | 4.85E+00 | 2.02E+00 | 4.27E+00 |
| 1000 | 1.30E+01 | 6.63E+00 | 3.43E+00 | 5.11E+00 |
| 1500 | 3.48E+01 | 2.23E+01 | 6.21E+00 | 9.81E+00 |
| 2000 | 4.48E+01 | 2.82E+01 | 5.48E+00 | 7.54E+00 |
| 2500 | 1.36E+02 | 8.74E+01 | 1.31E+01 | 4.01E+01 |
| 3000 | 1.03E+04 | 3.46E+03 | 2.97E+01 | 8.53E+02 |
| 3500 | 6.47E+05 | 2.30E+05 | 8.27E+01 | 2.23E+04 |
| 4000 | 4.09E+07 | 1.30E+07 | 1.69E+02 | 8.97E+05 |
| 4500 | 3.06E+08 | 7.27E+07 | 3.87E+02 | 5.45E+06 |
| 5000 | 1.23E+09 | 2.88E+08 | 8.32E+02 | 4.03E+07 |
| 5500 | 4.78E+09 | 1.15E+09 | 3.13E+03 | 4.73E+08 |
| 5643 | 5.09E+09 | 1.38E+09 | 3.10E+03 | 5.63E+08 |
| | 276.2% | 260.3% | 144.1% | 249.9% |

Table 3. LOLREC portfolio selection strategy with different voting functions compared to the best benchmark kernel-based log-optimal portfolio selection strategy reported in [2]

Further analysis shows that there is one asset KINAR, which was extremely predictable, thus generated a significant portion of the returns. The expert with average voting strategy of KINAR alone had $3.51*10^8$ return generated. This inevitable biased the results, which we will address in the next experiment. Note that by using individual predictors LOLREC also gives us the ability to find extremely predictable assets in a set of stocks that a trader can concentrate on.

We will see in later sections that the market structure changed. In recent years (ending in April 2012) the mode voting strategy is superior compared to the average and median voting strategy.

---

[3] Average annual yield is computed by the expression $S_i^{1/N}$, where N is the length of the dataset in years.

## 4.2. Results of LOLREC on NYSEMERGED dataset

We have tested the algorithm on NYSEMERGED that has a longer timeframe and the outliers like KINAR removed. As the benchmark papers report average annual yields, we will not do otherwise. We have reduced the number of committee members to 9: a committee $C_j([1,2,3],[1,2,3])$ was used with the three voting functions for each asset. This setup may reduce the AAY of each committee and the AAY of the portfolio, but it would have been computationaly infeasable to perform the experiment with a greater number of base experts. As for the previous experiment we will stick to selecting the top 10 performers (m=10).

We report that the performance measured by AAY is reduced to 26% for the average voting strategy, 22% for the median voting strategy and 19% for the mode voting strategy. This shows that the removal of outlier assets like KINAR (which were remarkably predictable) significantly reduced the AAY furthermore the reduction in committee members significantly impacted achieved AAY. The AAY reported in [12] for the kernel based semi-log-optimal portfolio is 31% for this timeframe. This shows that LOLREC performs best when extremely predictable assets are in the portfolio, although selecting a larger number of experts may increase the performance (this might be investigated in the future).

Figure 1 shows the wealth at each trading period.

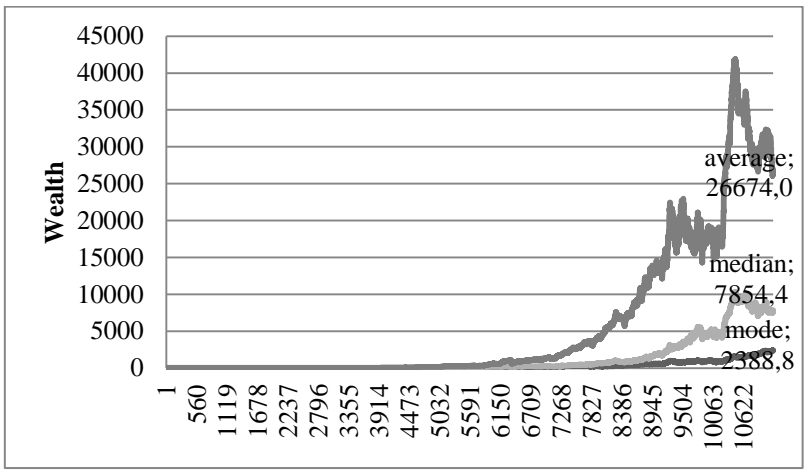

Figure 1. Wealth achieved by different voting strategies of LOLREC on NYSEMERGED dataset.

It is interesting to note that from the year 2003 both the average and median voting strategies lost a significant portion from its highest values, but the mode voting strategy had not been affected by the change in the market structure. To show how the change in the structure of the market affected the committees we perform an experiment on more recent data SNP500.

## 4.3. Results of LOLREC on SNP500

The SNP500 dataset has a lot more assets and also show how flexible LOLREC is. We test the portfolio diversification effect by changing the *m* parameter. We use committees $C_j([1,\ldots,10],[1,\ldots,10])$ so the next-day return is based on 100 base estimators for an individual asset. First we will show in Table 5 how the voting strategies performed on m=10.

|  | Mode | Median | Average | SNP500 buy-and-hold | Equally weighted portfolio |
| --- | --- | --- | --- | --- | --- |
| Minimum of capital | 0.8520 | 0.2698 | 0.2942 | 0.44818 | 0.4969 |
| Worst 1-period return | -17% | -16% | -15% | -10% | -10% |
| AAY | 37% | 10% | 7% | -2% | 0.1% |
| Standard deviation of 1-period returns | 0.029 |  |  | 0.017 | 0.019 |
| Average 1-period return | 1.002 |  |  | 1.0 | 1.0 |

Table 5. AAY, worst 1-period return and minimum of capital achieved by LOLREC selecting the top 10 performing stocks

Table 5 shows that the best voting strategy was the mode voting strategy, thus shed light on, that the market structure changed: the better performing voting strategies of the past are not as efficient in the future. If LOLREC strategy would be applied in today's market the mode voting strategy would be the optimal choice, with an AAY of 37%. The mode voting strategy did not lose a lot of value during trading: lost only 14% of its initial value, as opposed to the other voting strategies, where the average voting strategy lost 73%, the median voting strategy lost 71% of the initial capital. On the other hand the mode voting strategy had the worst 1-period return of -17%. The other strategies worst 1-day is also comparable to this value and not significantly better.

We will now perform experiments with the parameter *m* on the mode voting strategy since it was the most profitable. We will compare the results to the equally weighted portfolio and the buy-and-hold return of S&P500 index. After $m > 291$ LOLREC did not produce any different values meaning that the one time the maximum number of estimated next-period positive returns was 291. Figure 2 shows the returns of LOLREC as the function of *m* and the standard deviation of one-period returns. The optimum value is at *m*=9 (for the maximum return portfolio), although at *m*=16 and *m*=26 there are two local maxima offering less volatile returns.

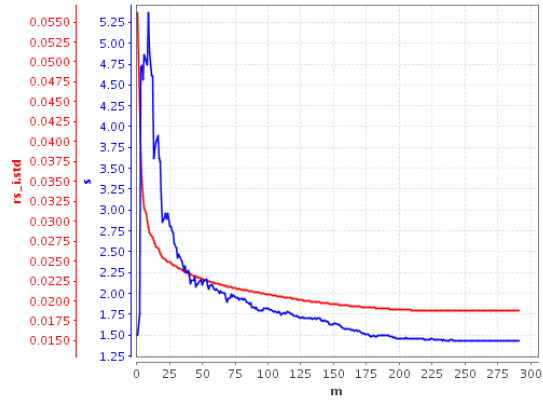

Figure 2. LOLREC cumulative returns and 1-period standard deviation as the function of m, the number of greatest estimated return assets

We show the resulting wealth gained from a hypothetical 1 unit investment in the first period for the LOLREC strategy compared to the equally weighted portfolio and the S&P500 index. Figure 3 shows that LOLREC significantly outperforms both baselines and achieves 5.3583 wealth in the end (this means 39% of AAY).

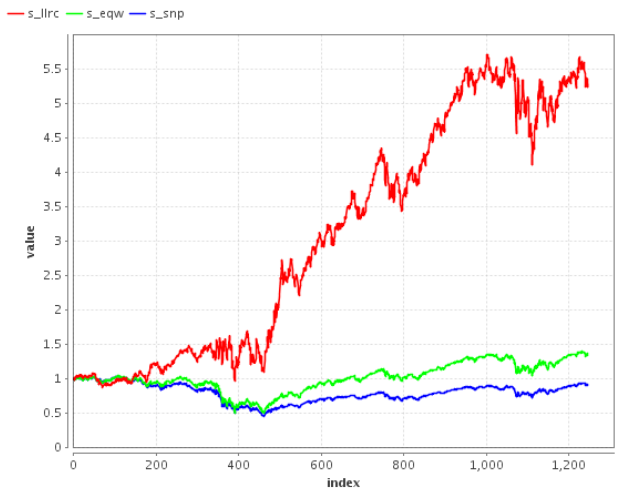

Figure 3. LOLREC portfolio weight selection yields and benchmark yields

By increasing the m parameter and selecting more stocks the wealth achieved converged to 1.4378 and std$(R_i)^4$ = 0.0187, which compared to the equal weighted portfolio's S=1.3627 and std$(R_i)$ = 0.0190 means that the algorithms not only produced greater final wealth on the most diversified version, it also produced less volatile returns or in other words a less risky portfolio than the equally weighted portfolio.

Analysis of the committees reveal the usefulness of the LOLREC strategy. On average the committees outperformed the buy-and-hold strategies by 40%. Table 6 sums up some findings about the committees.

|  | Ticker | Wealth of commitee ($S_C$) | Wealth of buy-and-hold ($S_{BNH}$) | Relative performance ($S_C/S_{BNH}$) | Average weight in portfolio vectors | No. times selected |
|---|---|---|---|---|---|---|
| Best committee | EQR | 7.33 | 1.676 | 4.374 | 0.253 | 43 |
| Worst committee | S | 0.087 | 0.127 | 0.021 | 0.021 | 58 |
| Best relative performance | C | 2.316 | 0.071 | 32.778 | 0.146 | 69 |
| Worst relative performance | HUM | 0.292 | 1.399 | 0.208 | 0.046 | 33 |
| Biggest average weight | HST | 5.842 | 0.725 | 8.056 | 0.287 | 49 |

Table 6. Data collected on individual committee performances on SNP500 dataset (m=9)

## 5. Future work

In this section we mention some possible future improvements of the algorithm. We have already mentioned that not only the committee's performances but each base expert's performance could be used when the voting takes place in the committee: this being a new kind of voting function, the performance weighted average vote. Further algorithm parameters like smoothing kernels and different distance functions could be examined to see if the overall yield grows or decreases. We have not demonstrated explicitly the effect of using more experts on yields however results not shown in this paper strongly suggest that more base experts increase the overall yield achieved by the portfolio selection strategy.

---

[4] standard deviation of the daily returns

# 6. Summary

In the paper we have shown that the proposed Local Linear Regression Ensemble Committee with heuristic weight selection based on past performance of the ensemble committees outperform a benchmark portfolio optimization technique that optimize the growth rate of the portfolio reported in [2]. Furthermore we showed practical relevance of the algorithm on recent real world data comprising of 453 different assets from the S&P500 index. With the mode voting function used in the committee an average annual yield of 39% percent can be reached with the selection of the top 9 assets with the largest estimated next-period return. If we increase the number of selected stocks the overall yield of the algorithm decreases, however every possible parameter setting outperforms the equally weighted portfolio's return and the S&P500's return on the timeframe both in terms of yield in the end and risk measured by the standard deviation of 1-period returns.